\documentstyle[epsf]{article}

\newcommand{\bea}{\begin{eqnarray}}
\newcommand{\eea}{\end{eqnarray}}
\newcommand{\beq}{\begin{equation}}
\newcommand{\eeq}{\end{equation}}
\newcommand{\nn}{\nonumber}
\newcommand{\marke}[1]{ \protect\label{#1}
}
\def\gsim{\mathrel{\raise.3ex\hbox{$>$\kern- .75em
\lower1ex\hbox{$\sim$}}}}

\newfont{\cms}{cmss8 scaled 1440}

\def\k{{\vec k}}
\def\x{{\vec x}}
\def\/{\over}
\begin{document}

\parindent=1 em
\frenchspacing

\noindent
\begin{minipage}[t]{7cm}\rule{70mm}{1mm} \\[2mm]
{\cms  Konstanz University \hfill $\bullet$  \hfill Theory Group
  \hfill \\[4mm]
$\bigl\langle$ \hfill Gravity \hfill $\bigr|$ \hfill Quantum
   Theory \hfill $\bigr|$ \hfill
   Optics \hfill $\bigr\rangle$} \hfill \\[1mm]
\rule{70mm}{1mm}
\end{minipage}
\hfill \\[0.75cm]
\noindent{\em Preprint  KONS-RGKU-95-07}\\[2cm]

\begin{center} {\LARGE \bf Generalized Unruh effect and Lamb shift
for atoms
\\[0.2cm]
on arbitrary stationary trajectories}
\\[1cm]
{\bf
J\"urgen Audretsch\footnote{e-mail: Juergen.Audretsch@uni-konstanz.de},
 Rainer M\"uller\footnote{Present address: Sektion Physik der
Universit\"at
 M\"unchen, Theresienstr. 37, D-80333 M\"unchen, Germany \\
\phantom{aaa} e-mail: Rainer.Mueller@physik.uni-muenchen.de}
 and Markus Holzmann
  \\[0.3cm]
\normalsize \it Fakult\"at f\"ur Physik der Universit\"at Konstanz\\
\normalsize \it Postfach 5560 M 674, D-78434 Konstanz, Germany}
\vspace{0.2cm}

\begin{minipage}{15cm}
\begin{abstract}
We study the spontaneous de-excitation and excitation of accelerated
atoms on arbitrary stationary trajectories (``generalized Unruh
effect''). We consider the effects of vacuum fluctuations and
radiation reaction separately. We show that radiation reaction
is generally not altered by stationary acceleration, whereas the
contribution
of vacuum fluctuations differs for all stationary accelerated
trajectories from its inertial value. Spontaneous excitation from
the ground
state
occurs for all { accelerated stationary} trajectories and is therefore
the ``normal case''. We furthermore show that the radiative energy
shift (``Lamb shift'') of a two-level atom is modified by
acceleration for all stationary trajectories. Again only vacuum
fluctuations give rise to the shift. Our results are illustrated
for the special case of an atom in circular motion, which may be
experimentally relevant.\\
PACS numbers: 32.80-t; 42.50-p.

\end{abstract}
\end{minipage}
\end{center}

\vspace{0.3cm}

\section{Introduction}

Recently, a new physical picture for the spontaneous excitation of
a uniformly accelerated two-level atom (the Unruh effect \cite{Unruh76})
has been put forward \cite{Audretsch94a}. Following a quantum
optical approach
{it} is based on the distinction of the two
competing mechanisms which {operate} to excite an atom in
the quantum vacuum: vacuum fluctuations and radiation reaction.
Vacuum fluctuations tend to excite an atom in its ground state and
de-excite it in an excited state. On the other hand, radiation
reaction leads always to a loss of internal excitation energy.
For an inertial atom in the ground state, the two contributions
cancel exactly, so that a very sublime balance between vacuum
fluctuations and radiation reaction prevents the spontaneous
excitation of the atom \cite{Dalibard82}. If the atom is in the
excited state, both contribution add up to the well-known spontaneous
emission rate. It has been shown in \cite{Audretsch94a} (see also
\cite{Sciama81}) that uniform acceleration will disturb this balance.
Spontaneous transitions from the ground state to the excited state
become possible: the Unruh effect. Similarly, the rate of spontaneous
emission is modified from its inertial value for a uniformly
accelerated atom.

It is interesting to investigate the ``fine structure'' of these
processes by considering vacuum fluctuations and radiation reaction
separately. In the case of uniform acceleration, the contribution
of vacuum fluctuations is changed from its inertial value, while
the contribution of radiation reaction remains exactly the same as
for the inertial atom. Their combined action leads to the spontaneous
excitation or de-excitation described above.

In this paper, we will concentrate on more general states of
motion of the atom. We consider arbitrary stationary trajectories in
Minkowski space, which can be characterized as a motion along a
timelike Killing vector field or, equivalently, by the condition
that the geodesic distance between two points on the trajectory
depends only on their proper time difference \cite{Letaw81}.
We will ask for which of these trajectories there is a spontaneous
excitation of the atom (``generalized Unruh effect''). Furthermore
we will discuss whether it is a general trait that the contribution
of radiation reaction is not modified by the acceleration or whether
this holds only in the special case of uniform acceleration.
Another question is whether it is
true also for more general trajectories that the atom is excited
because it perceives the vacuum fluctuations differently.

Our results are summarized in a number of theorems. The first one
shows that the contribution of radiation reaction is the same
for all stationary accelerated trajectories as in the inertial case.
Accordingly, radiation reaction is a purely local concept that is not
sensitive to the actual state of motion of the atom. This can be
understood heuristically: the field is radiated away by the atom
with the velocity of light. Since the atom's trajectory is timelike,
the only moment it can act back on the atom is precisely at the time
it is emitted. The mechanism of radiation reaction is thus restricted
to act at this point of the trajectory alone.

Next we show that the contribution of vacuum fluctuations differs
for all atoms on accelerated stationary trajectories from that of an
inertial atom. Together, these two results demonstrate that
a generalized Unruh effect takes place for all stationary trajectories
except the inertial ones. In this sense, the spontaneous excitation
of an atom is the normal case and the non-occurrence of the
Unruh effect for inertial atoms is an exception.

The second main part of the paper is concerned with the {radiative
energy shift}
of accelerated atoms. The  {``Lamb shift''} for a uniformly
accelerated two-level atom has been calculated
{already} in \cite{Audretsch94b}.
This shift can be explained qualitatively as follows: For a uniformly
accelerated atom, the Minkowski vacuum appears as a thermal bath
of particles called Rindler quanta. The AC Stark shift associated
with these particles is responsible for the atom's modified Lamb
shift.

In Sec. 4, we will investigate the Lamb shift for atoms on arbitrary
stationary trajectories, where a simple interpretation in terms of
Rindler particles is no longer possible. Already for an inertial
atom, the energy shift diverges and must be regularized, for example
by imposing a cutoff energy. Consequently, it may be difficult to
isolate the modifications introduced by the acceleration. We overcome
this problem by referring to a recently {proved} theorem
\cite{Audretsch95}
that connects the radiative energy shifts to the transition rates
between the levels, which are much easier to calculate. In this way,
we can show that the acceleration-induced modification of the Lamb
shift is always finite. This allows an unambiguous separation into
a (divergent) inertial part and a finite correction caused by the
acceleration.

We will then consider the individual contributions to the Lamb shift.
We will show that the effect of radiation reaction is the same
for an atom on an arbitrary stationary trajectory as for an inertial one.
It is equal for both levels and does not contribute to the relative
shift.
Contrarily, the effect of vacuum fluctuations is modified for any
accelerated stationary trajectory. The total shift differs always
from its
inertial value.

As an illustration of our results, we treat in Sec. 5 an atom in
circular motion. The spontaneous excitation for a two-level atom
on such a trajectory has been treated previously
\cite{Letaw80,Bell83,Takagi86} and may be of experimental relevance.
We discuss the contributions of vacuum fluctuations and radiation
reaction separately and determine the evolution of the mean atomic
excitation energy towards its equilibrium value. Furthermore,
we identify the Einstein coefficients for spontaneous excitation
and de-excitation. Finally, the Lamb shift for a circulating atom
is calculated by applying the theorem of Ref. \cite{Audretsch95}.
The correction due to the acceleration is isolated and compared with
the uniformly accelerated case.

\section{Vacuum fluctuations and radiation reaction: the general
formalism}

As a basis for the discussion of the behaviour of accelerated atoms in
arbitrary stationary motion in Special Relativity, we will briefly
recall
the model and the
Heisenberg picture formalism that was used to treat the case of uniformly
accelerated atoms in previous publications
\cite{Audretsch94a,Audretsch94b}.
We consider a pointlike two-level atom on an arbitrary stationary
space-time
trajectory $x(\tau)$ which interacts with a scalar quantum field $\phi$.
$\tau$ denotes the proper time on the trajectory. A characterization
of stationary motion in Minkowski space has been given by Letaw
\cite{Letaw81}, resulting in two equivalent conditions: (a) the
trajectory follows the orbits of a timelike Killing vector field, or
(b) the geodesic distance between two points $x(\tau)$ and
$x(\tau')$ on the trajectory depends only on the proper time interval
$\tau-\tau'$.

Property (a), mentioned above, ensures that the metric along the
path of the
atom does not change and guarantees the existence of stationary states.
For our two-level atom, these will be called $|+ \rangle$ and $|-
\rangle$, with energies $\pm {1\/ 2}\omega_0$. The atom's undisturbed
Hamiltonian with respect to its proper time $\tau$ can then be written
\beq H_A (\tau) =\omega_0 R_3 (\tau),  \marke{eqV1} \eeq
where $R_3 = {1\/2} |+ \rangle \langle + | - {1\/2}| - \rangle
\langle - |$
is the pseudospin operator commonly used in the description of
two-level atoms \cite{Dicke54,Allen75}.

The free Hamiltonian of the scalar quantum field that governs its
time evolution with respect to $\tau$ is
\beq H_F (\tau) = \int d^3 k\, \omega_\k \,a^\dagger_\k a_\k
	{dt\/d \tau} \marke{eqV3}\eeq
with creation and annihilation operators $a^\dagger_\k$,  $a_\k$.
Atom and field are coupled by the interaction
\beq H_I (\tau) = \mu R_2 (\tau) \phi ( x(\tau)) \marke{eqV5}\eeq
where $\mu$ is a small coupling constant , $R_2 = {1\/2} i ( R_- -
R_+)$,
and $R_+ = |+ \rangle \langle - |$, $R_- = |- \rangle \langle +|$.
Note that the atom-field coupling (\ref{eqV5}) is effective only on
the trajectory of the atom.

We can now write down the Heisenberg equations of motion for the atom and
field observables.The field is always considered to be in its special
relativistic vacuum state $|0 \rangle$  ({ Minkowski vacuum}).
To isolate the contributions of vacuum fluctuations
and radiation reaction to the rate of change of atomic observables,
their solution for the field $\phi$ is split into its free or vacuum
part $\phi^f$, which is present even in the absence of the coupling,
and its source part $\phi^s$, which represents the field generated by
the atom itself. If we consider now the effects of $\phi^f$ and $\phi^s$
separately in the Heisenberg equations of an arbitrary atomic
observable $G$, we obtain the {individual}
contributions of {\it vacuum fluctuations}
and {\it radiation reaction} to the rate of change of $G$. Following
Dalibard, Dupont-Roc and cohen-Tannoudji \cite{Dalibard82}, we choose a
symmetric ordering between atom
and field variables. Below we repeat only the necessary definitions and
central expressions of \cite{Audretsch94a} and \cite{Audretsch94b}.

Our first aim is the discussion of spontaneous emission and spontaneous
excitation from the ground state $|- \rangle$ which we will call {\it
generalized Unruh
effect}. We therefore concentrate on the mean atomic excitation
energy $\langle H_A (\tau) \rangle$. It has been shown in
\cite{Audretsch94a} that the contributions of vacuum fluctuations
(vf) and
radiation reaction (rr) to the rate of change of $\langle H_A \rangle$
can be written (cf. \cite{Dalibard82,Dalibard84})
\bea \left\langle {d H_A (\tau) \/ d\tau} \right\rangle_{vf} &=&
	2 i \mu^2  \int_{\tau_0}^\tau d \tau' \, C^F(x(\tau),x(\tau'))
	{d\/ d \tau} \chi^A(\tau,\tau'), \marke{eqV26}\\
\left\langle {d H_A (\tau) \/ d\tau} \right\rangle_{rr} &=&
	2 i \mu^2
	\int_{\tau_0}^\tau d \tau' \, \chi^F(x(\tau),x(\tau')) {d\/
d \tau}
	C^A(\tau,\tau'). \marke{eqV27}\eea
with $| \rangle = |a,0 \rangle$ representing the atom in the state
$|a\rangle$ and the field in the Minkowski vacuum state $|0
\rangle$.They are
expressed in terms of the statistical functions of the atom
\bea C^{A}(\tau,\tau') &:=&  {1\/2} \langle a| \{ R_2^f (\tau),
R_2^f (\tau')\}
	| a \rangle \marke{eqV28} \\
\chi^A(\tau,\tau') &:=& {1\/2} \langle a| [ R_2^f (\tau), R_2^f (\tau')]
	| a \rangle. \marke{eqV29}\eea
and of the field
\beq C^{F}(x(\tau),x(\tau')) := {1\/2} \langle 0| \{ \phi^f
(x(\tau)), \phi^f
	(x(\tau')) \} | 0 \rangle, \marke{eqV24} \eeq
\beq \chi^F(x(\tau),x(\tau')) := {1\/2} \langle 0| [ \phi^f
(x(\tau)), \phi^f
	(x(\tau'))] | 0 \rangle. \marke{eqV25}\eeq
$C^A$ and $C^F$ are called symmetric correlation functions, $\chi^A$ and
$\chi^F$ linear susceptibilities. Note that the dependence on the
trajectory $x(\tau)$ of the atom in (\ref{eqV26}) and (\ref{eqV27}) is
contained
entirely in the statistical functions of the field, which have to be
evaluated at two points on the atom's world line. The atom registers the
influence of the Minkowski vacuum when following its
specific space-time trajectory.

The explicit form of the statistical functions of the atom is given by
\bea C^{A}(\tau,\tau') & =& {1\/2} \sum_b|\langle a | R_2^f (0) | b
	\rangle |^2 \left( e^{i \omega_{ab}(\tau - \tau')} + e^{-i
\omega_{ab}
	(\tau - \tau')} \right), \marke{eqV30}\\
\chi^A(\tau,\tau') & =& {1\/2}\sum_b |\langle a | R_2^f (0) | b
\rangle |^2
	\left(e^{i \omega_{ab}(\tau - \tau')} - e^{-i \omega_{ab}(\tau -
\tau')}
	\right), \marke{eqV31}\eea
where $\omega_{ab}= \omega_a-\omega_b$ and the sum extends over a
complete
set of atomic states.
The statistical functions of the field are
\bea C^{F}(x(\tau),x(\tau')) & =& {1\/8 \pi^2} {1\/|\Delta \x |} \left(
	{{\cal P}\/ \Delta t + |\Delta \x|} - {{\cal P}\/ \Delta t-
	|\Delta \x|}
	\right) \nonumber\\
	&=& - {1\/ 8\pi^2} \left( {1 \/(\Delta t+ i \epsilon)^2 -
	|\Delta \x|^2} +
	{1 \/(\Delta t- i \epsilon)^2 - |\Delta \x|^2}\right),
\marke{eqV32}\\
\chi^F(x(\tau),x(\tau')) & =& { i\/ 8 \pi}{1\/ |\Delta \x|} \left(
\delta(
	\Delta t + \Delta|\x|)- \delta(\Delta t - |\Delta \x|) \right),
	\nonumber\\
	&=& {1\/ 8\pi^2} \left( {1 \/(\Delta t+ i \epsilon)^2 -
	|\Delta \x|^2} -
	{1 \/(\Delta t- i \epsilon)^2 - |\Delta \x|^2}\right),
\marke{eqV33a}
	\eea
where $\Delta t=t(\tau) - t(\tau')$, $\Delta \x = \x (\tau) - \x (\tau')$
and $\cal P$ denotes the principal value.

In a next step, we consider the ``Lamb shift'' of the two-level
atom, i. e.
the radiative energy shifts of its levels due to the coupling to
the quantum vacuum. To this end, one studies the evolution of an
atomic observable $G$ and traces out the field degrees of freedom
in its equations of motion. It is then possible to identify an
effective Hamiltonian for the atom which acts in addition to $H_A$
(cf. \cite{Dalibard82,Dalibard84}). The expectation value of $H_{eff}$
in an atomic state $|a \rangle$ gives the radiative shift of this
level. Again the division of the field into free and source part
allows the identification of the contributions of vacuum fluctuations
and radiation reaction. As has been shown in \cite{Audretsch94b},
the radiative energy shifts of level $|a \rangle$ are given by
\beq (\delta E_a)_{vf} = - i \mu^2 \int_{\tau_0}^\tau d \tau' \,
	C^F(x(\tau),x(\tau')) \chi^A(\tau,\tau'), \marke{eqS34} \eeq
\beq (\delta E_a)_{rr} = - i \mu^2
	\int_{\tau_0}^\tau d \tau' \, \chi^F(x(\tau),x(\tau'))
	C^A(\tau,\tau'). \marke{eqS35}\eeq

As we restrict ourselves to the model of a two level atom the
statistical
functions of the atom (\ref{eqV30}) and (\ref{eqV31}) can be
simplified: The
level spacing $\omega_{ab}$ is given by $|\omega_{ab}|=\omega_o
\delta_{a,b}$
 and the matrix element $|\langle a | R_2^f (0) | b \rangle |^2$ is
reduced to
$$|\langle a | R_2^f (0) | b \rangle |^2 = {1 \over 4}(1 -
\delta_{a,b}).$$
Hence the summation in (\ref{eqV30}) and (\ref{eqV31}) breaks down.
In the following theorems we nevertheless use the unevaluated formulas
(\ref{eqV30}) and (\ref{eqV31}) to stress the structure of the
results so that possible extensions to multi-level atoms are easily
made (cf. \cite{Audretsch95}).

\section{Relaxation rates for atoms on arbitrary
stationary trajectories}

Based on the formalism presented above, we will discuss in this
section some
general features of the relaxation rates (\ref{eqV26}) and (\ref{eqV27})
for atoms in arbitrary stationary motion. Two special cases, discussed in
a previous paper \cite{Audretsch94a}, are an unaccelerated atom in
uniform
motion (or at rest) and a uniformly accelerated atom with constant
acceleration $a$.
It was found in \cite{Audretsch94a} that for a uniformly accelerated atom
the contribution of radiation reaction (\ref{eq5}) to $\langle dH_A
/d \tau \rangle$ is the same as for an atom at rest. On the other hand,
the contribution of vacuum fluctuations is changed by the influence of
acceleration. This means that spontaneous excitation of a
uniformly accelerated
atom in the ground state becomes possible and provides a physical
picture of the Unruh effect.

The question arises: is this structure a special feature of the uniformly
accelerated trajectory or does it
appear in a similar way for other stationary trajectories too? We will
prove two
theorems which show that indeed spontaneous excitation of the atom
takes place for any non-inertial trajectory

\smallskip
{\it Theorem 1: The contribution of radiation reaction to the rate of
change $\langle {dH_A \over {d \tau}} \rangle$ of the mean atomic energy
is for all stationary trajectories equal to that of an atom at rest:}
\beq \left\langle {d H_A (\tau) \over  d\tau} \right\rangle_{rr}
	= \sum_b \Gamma_{ab}^{rr} (\omega_{ab}), \marke{eqM1}\eeq
where
\beq \Gamma_{ab}^{rr} (\omega):=
	-{\mu^2 \over  4\pi} \omega^2 |\langle a | R_2^f (0)
	|b\rangle |^2. \marke{eq868}\eeq

\smallskip

{\it Proof:} In Eq. (\ref{eqV27}), the only dependence on the trajectory
 of the atom is through the linear susceptibility of the field. It is
therefore sufficient to analyze $\chi^F$. We will show that it is equal
in the distributional sense (i. e. integrated over an arbitrary test
function) for all timelike trajectories. Starting from Eq.
(\ref{eqV33a}),
we apply $\chi^F$ to a test function $f(\tau')$.
Since its Fourier transform $\tilde f(\omega)$ exists, it can be split in
a part $f^{(+)}(\tau')$ that vanishes for $\tau'\to +i\infty$ and a
part $f^{(-)}(\tau')$ that vanishes for $\tau'\to -i\infty$:
$$ f(\tau') = f^{(+)}(\tau') + f^{(-)}(\tau') :=
  \int_0^\infty d\omega \left( \tilde f(\omega) e^{i\omega \tau'}
  + \tilde f(-\omega) e^{-i\omega \tau'} \right) $$
We first treat the $f^{(+)}$ part of the integral:
\beq \int d\tau' f^{(+)}(\tau') \chi^F(x(\tau), x(\tau')) = {1\/ 8\pi^2}
	\int d\tau' f^{(+)}(\tau')  \left( {1 \/(\Delta t+ i
\epsilon)^2 -
	|\Delta \x|^2} -{1 \/(\Delta t- i \epsilon)^2 - |\Delta
\x|^2}\right)
	\label{eq*}\eeq
We calculate the integral via the residue theorem and treat first the
case where the contour can be closed in the upper half complex $\tau'$
plane. Due to the $i \epsilon$ structure of the two parts of $\chi^F$,
only the poles on the real $\tau'$ axis contribute. We notice that  for
real $\tau'$, the denominator of (\ref{eq*}) vanishes only for
lightlike separated points $x(\tau)$ and $x(\tau')$. Because our
trajectory
is timelike, the only contribution comes from $\tau'=\tau$. We can
therefore expand the denominator in (\ref{eq*}) around this point:
\beq (\Delta t \pm i\epsilon)^2 -|\Delta \vec x|^2 \approx
	(\tau-\tau')^2\left[\left({dt(\tau')\/ d\tau'}\right)^2 -
	\left({d\vec x (\tau')\/ d\tau'}\right)^2\right]_{\tau'=\tau}
	\pm 2 i \epsilon \left({dt(\tau')\/ d\tau'}\right)_{\tau'=\tau}.
	\eeq
The term in square brackets is just the square of the four-velocity
$u^\mu u_\mu=1$ and we obtain
\beq (\Delta t \pm i\epsilon)^2 -|\Delta \vec x|^2 \approx
	(\tau-\tau' \pm i\epsilon )^2 \eeq
where we have absorbed the positive quantity $dt/d\tau'$ in $\epsilon$.

The evaluation of the integral (\ref{eq*}) is now straightforward.
One finds
\beq \int d\tau' f^{(+)}(\tau') \chi^F(x(\tau), x(\tau')) = {i\/ 4\pi}
	\int d\tau' f^{(+)}(\tau')\, \delta'(\tau -\tau').
\label{eq**}\eeq
For the $f^{(-)}$ part which vanishes for $\tau'\to -i\infty$
the contour must be closed in the lower half plane and the same result
is found. By adding the two contributions, the desired result is
obtained.

Eq. (\ref{eq**}) shows that for any accelerated trajectory, $\chi^F$
has the same functional form as for an inertial one. Note that it was
not necessary to assume stationarity in deriving (\ref{eq**}).

To complete the proof, we calculate from (\ref{eqV27}) and (\ref{eqV30})
the contribution of radiation reaction to the rate of change of
$\langle H_A \rangle$:
\beq \left\langle {d H_A (\tau) \over  d\tau} \right\rangle_{rr} =
	-2 i \mu^2 \sum_b \omega_{ab} |\langle a | R_2^f (0)
|b\rangle |^2
	\int_{0}^{+\infty} du\, \chi^F(u) \sin\omega_{ab}u
\marke{eq866}\eeq
where $u=\tau-\tau'$. The integrand is symmetric so that we can
extend the lower limit to $-\infty$. The evaluation of the integral
is trivial and we obtain (\ref{eqM1}) with (\ref{eq868}).
This expression is independent of the trajectory $x(\tau)$.
It is the same formula
we found already in \cite{Audretsch94a} for an inertial atom and a
uniformly accelerated atom.

It is worth noting, that the structure of (\ref{eqM1}) and
(\ref{eq868}) is
{maintained} for multi-level atoms, since we did not use any properties
restricted to the two-level model.
\smallskip

{\it Theorem 2: The contribution of vacuum fluctuations to $\langle
dH_A/d \tau \rangle$ differs for any accelerated
stationary trajectory from that of an atom at rest.}
\smallskip

{\it Proof:} Using (\ref{eqV31}), we write Eq. (\ref{eqV26})
in the form
\beq \left\langle {d H_A (\tau) \over  d\tau} \right\rangle_{vf} =
	\sum_b \Gamma_{ab}^{vf} (\omega_{ab}) \marke{eq869}\eeq
with
\beq \Gamma_{ab}^{vf} (\omega):=
	 -2\mu^2 \omega |\langle a | R_2^f (0) |b\rangle |^2
	 \int_0^\infty du\, C^F(u) \cos\omega u. \marke{eq869a}\eeq
We recognize that the integral appearing in this expression is the
Fourier cosine transform of the symmetric correlation function
$C^F(u)$. Hence, to show that (\ref{eq869}) deviates from its inertial
value for a noninertial form of $C^F(u)$, we can use the uniqueness
theorem of the Fourier transformation: if $C^F(u)$ is of bounded
variation and the integral converges, the Fourier integral is unique.
The first requirement is trivially fulfilled for the monotonic
function $C^F(u)$. To show that the second one is also met, we use
the explicit form (\ref{eqV32}) of the symmetric correlation function.
Apart from the convergence factor, the term appearing in the
denominator is the geodesic distance $\sigma (\tau-\tau')$ between the
two points $x(\tau)$ and $x(\tau')$ on the trajectory of the atom,
expressed
as a function of the proper time $\tau$. Now we use the fact that for
any two points, the four-distance along their connecting geodesic
is maximal. Therefore
\beq \sigma (\tau,\tau') >|\tau-\tau'|. \eeq
Here
\beq \tau-\tau' = \int_{\tau'}^\tau d\tilde\tau\eeq
is the four-distance as measured along the world line. Hence we
always have
\beq {1\over  \sigma^2(\tau,\tau')} < {1\over (\tau-\tau')^2}
	\marke{eq872}\eeq
and the Fourier integral exists.

But the right-hand side of (\ref{eq872}) is just the expression
appearing in
the symmetric correlation function for an inertial atom (cf. Eq. (35)
of \cite{Audretsch94a}). Thus we have
\beq C^F (x(\tau),x(\tau')) <
	C^F_{inert} (x(\tau),x(\tau')) \marke{eq873}\eeq
for any two $\tau,\tau'$. From the uniqueness of the Fourier
transform we
now can say that
\beq \Gamma_{ab}(\omega) \not= \Gamma_{ab}^{inert}(\omega)
\marke{eqM3} \eeq
for any accelerated stationary  atom. Using (\ref{eq869}), the proof
for a two level atom is complete.

\smallskip
{\it Consequence:}
Taking the contents of Theorems 1 and 2 together, we see that for an
accelerated atom in its ground state, the balance between vacuum
fluctuations and radiation reaction, which prevents spontaneous
excitations for inertial atoms, is never maintained in the non-inertial
case. Spontaneous transitions
from the ground state to the excited state therefore take place on an
arbitrary accelerated stationary trajectory, leading to the generalized
Unruh effect. In addition,
the spontaneous emission rate for an atom in the excited state
is always altered through the influence of the {not necessarily uniform}
acceleration.

\section{Lamb shift for arbitrary stationary trajectories}

As mentioned in Sec. 2, it is possible with the same formalism to
identify the contributions of vacuum fluctuations and radiation
reaction to the Lamb shift of the two-level atom. It is interesting
to ask the same questions for the energy shifts as for the relaxation
rates: how are the two contributions $(\delta E_a)_{vf}$ and
$(\delta E_a)_{rr}$ of (\ref{eqS34}) and (\ref{eqS35}) modified for an
atom on an arbitrary stationary trajectory?

To investigate this problem we do not start directly from the formulas
 (\ref{eqS34}) and (\ref{eqS35}). Instead we use a relation between
relaxation rates and energy shifts that was derived under very general
assumptions in our previous paper \cite{Audretsch95}. Applied to the
special model we are discussing here, it reads
\beq \left( \delta E_a \right) _{rr} = - {1 \over {4 \pi}} \sum_b
	\int_{-\infty}^{+\infty} \,d\omega' {1 \over {\omega'}}
	\Gamma_{ab}^{rr}(\omega') \left( {{\cal P} \over \omega' +
	\omega_{ab}}+{{\cal P} \over \omega' - \omega_{ab}} \right).
	\label{eq883}\eeq
\beq \left( \delta E_a \right) _{vf} = - {1 \over {4 \pi}} \sum_b
	\int_{-\infty}^{+\infty} \,d\omega' {1 \over {\omega'}}
	\Gamma_{ab}^{vf}(\omega') \left( {{\cal P} \over \omega' +
	\omega_{ab}}-{{\cal P} \over \omega' - \omega_{ab}} \right),
	\label{eq882}\eeq
where $\Gamma_{ab}^{vf}$ and $\Gamma_{ab}^{rr}$ have been defined in
(\ref{eq869a}) and (\ref{eq868}). With the help of this theorem, it
is easy
to formulate the analogue of Theorem 1 for the radiative energy
shifts, especially for the relative shift $\Delta= \delta E_+
-\delta E_-$
of the two levels (``Lamb shift''), which is the measurable quantity:
\smallskip

{\it Theorem 3: (a) The contribution of radiation reaction to the energy
shift $\delta E_a$ is for all stationary trajectories equal to
that of an atom at rest. (b) Radiation reaction does not contribute
to the relative energy shift of a two-level atom:
\beq \Delta_{rr}=(\delta E_+)_{rr} - (\delta E_-)_{rr} = 0.
\label{eq32}\eeq
}

\smallskip
{\it Proof: } Both conclusions follow simply from Eq. (\ref{eq883})
together with (\ref{eq868}).

Since equation (\ref{eq883}) is not restricted to the two-level atom,
part (a) of Theorem 3 can be extended to multi-level atoms.
Part (b) will not remain true in the general case because of
different coupling strengths of the energy levels to the
field. However, there will be no influence of the motion to the relative
shift and it will remain constant.

It is not so easy to obtain an analogue of Theorem 2, because of the
divergence of the level shift of the atom at rest. We must first show
that the additional level shift to the inertial one due to the influence
of the motion is finite.
\smallskip

{\it Theorem 4: The additional contribution $(\delta E_a)_{vf} -
(\delta E_a)_{vf}^{inert}$ of vacuum fluctuations to
$\delta E_a$ due to {acceleration} is finite for all stationary
trajectories.}

\smallskip
{\it Proof:} We want to show that
\bea && |(\delta E_a)_{vf} - (\delta E_a)_{vf}^{inert}| \marke{eqM2}\\
  && =\left| - {1 \over
{4\pi}} \sum_b \int_{-\infty}^{+\infty} d\omega' \, \left(
{\Gamma_{ab}^{vf}
(\omega') \over {\omega'}} - {\Gamma_{ab}^{vf}(\omega')^I \over
{\omega'}}
\right) \left( {{\cal P} \over
{\omega' + \omega_{ab}}} -  {{\cal P} \over {\omega' -
\omega_{ab}}} \right)
\right| < \infty \nonumber \eea
{We estimate} the integrand for the case  $|\omega'|
\rightarrow \infty$. Using equation (\ref{eq869a}) with
(\ref{eqV32}) the
integrand is proportional to
\beq \left|{1 \over {\omega'^2 - \omega_{ab}^2}} \,
\int_0^{\infty} du \,
\left( {1 \over {\sigma^2(u+ i\epsilon) }} + {1 \over {\sigma^2(u -
i\epsilon)
}} - {1 \over {u^2}+i\epsilon} - {1 \over {u^2} - i\epsilon}\right) \cos
(\omega'u)\, \right| \eeq
with the abbreviation $\sigma^2(u+ i\epsilon):= (\Delta t+ i
\epsilon)^2 -
	|\Delta \x|^2$. Because of the stationary motion the right side
depends only on the proper time interval $u=\tau-\tau'$
($\sigma^2(u)=u^2$
for an inertial atom).
With the substitution $x =\omega' u$ we can expand
\beq {1 \over {\sigma^2(x)}}={\omega'^2 \over {x^2(1 + A{x^2 \over
{\omega'^2}} +{\cal O}( {x^4 \over {\omega'^4}}))}} \approx {\omega'^2
\over {x^2}} (1- A{x^2 \over {\omega'^2}} +{\cal O}( {x^4 \over
{\omega'^4}})) \eeq
 in the limit  $|\omega'| \rightarrow \infty$.
A short calculation shows that the integrand of (\ref{eqM2}) goes
faster to
zero as ${ 1\over {\omega'^2}}$ and hence the total integral converges.

This fact agrees with our expectation that the additional energy
shift due
to an accelerated motion should be a continuous function of the
acceleration without divergences, so that for weak acceleration this
additional contribution goes continuously to zero.
\smallskip

{\it Theorem 5: The contribution of vacuum fluctuations to $\delta E_a$
differs for any accelerated
stationary trajectory from that of an atom at rest. }

\smallskip
{\it Proof:} Using equation (\ref{eqS34}) with the linear
susceptibility of
the atom (\ref{eqV31}) we can write for the additional contribution
to the
inertial value of the energy shift
\beq (\delta E_a)_{vf} - (\delta E_a)_{vf}^{inert}=  \mu^2 |\langle
a | R_2^f
(0)
	|b\rangle |^2 \int_0^{\infty}du \, [C^F(u)-C^F_{inert}(u)]
\sin\omega_{ab}u. \eeq
The integral appearing is just the Fourier sine transform of $C^F(u)-
C^F_{inert}(u)$. In Theorem 4 we have shown that the integral converges,
so that the Fourier transform exists and is unique. Equation
(\ref{eq873})
ensures that there will always be a
contribution of vacuum fluctuations to the energy shift of an atom at
rest for any accelerated stationary motion.

It is easy to see, that for the two level system the relative
energy shift
is given by
\beq \Delta_{vf}=(\delta E_+)_{vf} - (\delta E_-)_{vf}=2\,(\delta
E_+)_{vf}.
\eeq
This agrees with the total Lamb shift $\Delta$ for a two-level
atom, since
according to (\ref{eq32}), radiation reaction
does not contribute to the relative shift of
the two levels. Because of the divergence of the Lamb shift,
it is useful to split the
Lamb shift in the diverging part $\Delta_{inert}$, which is equal to the
Lamb shift of an atom at rest, and the non-diverging rest ${\cal D} $
appearing for
any accelerated moving atom
\beq \Delta = \Delta_{inert} + {\cal  D}.\eeq
Below we will evaluate $\cal D$ for a special case.

{\it Result:} The Lamb shift $\Delta$ of an accelerated
two-level atom always deviates
from its inertial value, although only the contribution of vacuum
fluctuations gives rise to this modification.

\section{Application: Atom in circular motion}

To illustrate the use of our theorems, we consider the concrete
situation of an atom in circular motion on the trajectory
\beq t(\tau)=\gamma\tau, \qquad x(\tau)=R \cos(\omega\gamma\tau) ,
\marke{eq1}\eeq
$$ y(\tau)=R \sin(\omega\gamma\tau), \qquad z(\tau)=0$$
with constant radius $R$ and magnitude of velocity $v$. In (\ref{eq1}),
$\gamma=(1-v^2)^{-1/2}$ is the usual relativistic factor, $\omega=
v/R$ is the angular velocity, and $a=v^2\gamma^2/R=\omega^2\gamma^2 R$
the magnitude of acceleration. The ``circular Unruh effect''
for a two-level system on the trajectory (\ref{eq1}) has been discussed
by several authors \cite{Letaw81,Bell83,Takagi86}. We will discuss
separately the contributions of vacuum fluctuations and radiation
reaction to
the Unruh effect and spontaneous emission. Furthermore, we will be able
to study the evolution of the atomic population to its equilibrium
value and identify the Einstein coefficients for the spontaneous
processes. Finally, application of Eqs. (\ref{eq882}) and (\ref{eq883})
will allow us to
determine the Lamb shift of an atom in circular motion.

The statistical functions of the field can be evaluated with the help of
\beq |{\bf x}(\tau)-{\bf x}(\tau')|=2R \left|\sin \left(\omega\gamma{
\tau-\tau' \over 2}\right)\right|. \marke{eq3}\eeq
We find
\beq \chi^F\left({\bf x}(\tau),{\bf x}(\tau')\right)=-{i \over {8\pi}}
	{\omega \gamma \over {|\sin(\omega \gamma {\tau-\tau' \over
	2})|}}\delta(\tau-\tau') \marke{eq4}\eeq
\bea C^F\left({\bf x}(\tau),{\bf x}(\tau')\right)&=&-{1 \over {8 \pi^2}}
	\Biggl({1 \over {\gamma^2(\tau-\tau')^2-4R^2\sin^2(\omega \gamma
	{\tau-\tau' \over 2})+i\epsilon}}\marke{eq5}\\
	&&\qquad\qquad + {1 \over{\gamma^2(\tau-\tau')^2
	-4R^2\sin^2(\omega \gamma {\tau-\tau' \over
2})-i\epsilon}}\Biggr)
	\nonumber\eea
It is easy to verify by explicit calculation that the contribution of
radiation reaction to $\langle dH_A/d \tau \rangle$ is given indeed
by (\ref{eq868}), as stated by Theorem 1.

The contribution of vacuum fluctuations is more difficult to obtain.
It is calculated from Eq. (\ref{eqV26}) via the residue theorem so that
we need to know the zeroes of the denominator. This leads to an
equation of the form $x^2=v^2 \sin^2 x$, which is not analytically
solvable. The problem becomes tractable in the ``high velocity limit''
\cite{Bell83,Takagi86}. For $v \gsim 0.85$, we can expand the sine
to find the zero with the smallest imaginary part (besides $x=0$).
Because the exponential in the numerator of the resulting integral,
zeroes with larger imaginary part can be neglected. In this way,
we obtain
\bea \left\langle { dH(\tau) \over {d\tau}} \right\rangle_{vf}&=&
	-{\mu^2 \over {2\pi}} \sum_{\omega_a>\omega_b} \left(
	{\omega_{ab}^2 \over 2} + {a \omega_{ab} \over {4 \sqrt{3}}}A
	e^{-2 \sqrt{3} B {\omega_{ab}\over a}} \right)  |\langle
	a|R_2^f(0)|b\rangle|^2 \marke{eq7}\\
	&&\qquad - {\mu^2 \over {2\pi}}\sum_{\omega_a<\omega_b}
	 \left( {\omega_{ab}^2 \over 2}
	+ {a |\omega_{ab}| \over {4 \sqrt{3}}}A e^{-2 \sqrt{3} B
	{|\omega_{ab}
	|\over a}} \right)  |\langle a|R_2^f(0)|b\rangle|^2\nn\eea
where
\beq A=1+{3 \over 5}(v\gamma)^{-2}, \qquad
	B=1-{1 \over 5}(v\gamma)^{-2}. \marke{eq8}\eeq
The total rate of change of the atomic excitation energy can be found
by adding the contributions of vacuum fluctuations and radiation
reaction:
\bea \left\langle { dH(\tau) \over {d\tau}} \right\rangle_{tot}&=&
	-{\mu^2 \over {2\pi}}\sum_{\omega_a>\omega_b} \left(\omega_{ab}^2
	+ {a \omega_{ab} \over {4 \sqrt{3}}}A e^{-2 \sqrt{3} B
{\omega_{ab}
	\over a}} \right)  |\langle a|R_2^f(0)|b\rangle|^2 \marke{eq9}\\
	&&\qquad -{\mu^2 \over {2\pi}} \sum_{\omega_a<\omega_b}
	{a |\omega_{ab}| \over {4
	\sqrt{3}}}A e^{2 \sqrt{3} B {|\omega_{ab}|\over a}}|\langle
	a|R_2^f(0)|b\rangle|^2 \nn\eea
We recognize that spontaneous excitation ($\omega_a<\omega_b$)
is possible as well as spontaneous de-excitation ($\omega_a>\omega_b$).
As stated by the contents of theorems 1 and 2, the balance between
vacuum fluctuations and radiation reaction which is present for an
inertial atom becomes upset.

{}From Eq. (\ref{eq9}), it is possible to get in second order in $\mu$
a differential equation for $\langle H_A \rangle$, (cf.
\cite{Audretsch94a})
\beq \left\langle { dH(\tau) \over {d\tau}} \right\rangle =-{\mu^2 \over
	{8\pi}}\omega_0 \left\{ {\omega_0 \over 2} + \left( 1 + {A
\over {2
	\sqrt{3}}} {a \over \omega_0} e^{-2\sqrt{3}B{\omega_0 \over a}}
	\right) \left\langle H(\tau) \right\rangle \right\}.
\marke{eq13}\eeq
The solution
\bea \left\langle H(\tau) \right\rangle &=& - {\omega_0 \over 2} +
	{\omega_0 \over 2}\left[1+{2 \sqrt{3} \over A} {\omega_0 \over a
	}e^{2\sqrt{3}B {\omega_0 \over a}}\right]^{-1} \nonumber\\
	&&\quad + \left( \left\langle H(0)
	\right\rangle +{\omega_0 \over 2} - {\omega_0 \over 2}
\left[1 + {2
	\sqrt{3} \over A} {\omega_0 \over a }e^{2\sqrt{3}B
{\omega_0 \over
	a}}\right]^{-1} \right) e^{-\Gamma \tau} \marke{eq14}\eea
gives the time evolution of the mean atomic excitation energy. {
The decay rate is}
\beq \Gamma = {\mu^2 \over {8\pi}}\omega_0 \left(1 + {A \over
{2\sqrt{3}}}
{a \over \omega_0 }e^{-2\sqrt{3}B {\omega_0 \over a}}\right).
\marke{eq15}\eeq
The second term in the bracket represents the modification of the
inertial
decay constant
\beq \Gamma_{inert}={\mu^2 \over {8\pi}}\omega_0. \eeq

{}From (\ref{eq14}) we see that the system evolves towards an equilibrium
population
\beq \langle H_A \rangle = - {1 \over 2}\omega_0 +
	{\omega_0 \over 2} \left(1 + {2 \sqrt{3} \over A} {\omega_0
\over a
	}e^{2\sqrt{3}B {\omega_0 \over a}}\right)^{-1}, \marke{eq17}\eeq
representing a mean excitation above the ground state $-{1\/ 2}
\omega_0$.

Comparison of (\ref{eq14}) with the solution of an appropriately defined
rate equation (cf. Eq. (65) in Ref. \cite{Audretsch94a}) allows
the identification of the two Einstein coefficients $A_\downarrow$
and $A_\uparrow$ for spontaneous emission and the Unruh effect.
We obtain
\beq A_{\downarrow}={\mu^2 \over {8\pi}}\omega_0 \left[1 + {A \over
	{4\sqrt{3}}} {a \over \omega_0 }e^{-2\sqrt{3}B {\omega_0
\over a}}
	\right], \marke{eq16a}\eeq
\beq A_{\uparrow} = {\mu^2 \over {8\pi}}a{A \over {4\sqrt{3}}}
	e^{-2\sqrt{3}B {\omega_0 \over a}}. \marke{eq16b}\eeq
The spontaneous emission rate $A_\downarrow$ shows a modification
of the value $\mu^2 \omega_0 /8\pi$ for an atom at rest, in
accordance with
Theorem 2. Spontaneous excitation (the ``circular Unruh effect'')
occurs with a rate $A_\uparrow$. Its value is in accordance with
previous results \cite{Bell83,Takagi86},

Following \cite{Bell83} we can define an {\it effective temperature} by
\beq kT_{eff} = \omega_{0} [- ln(A_{\uparrow}/A_{\downarrow})]^{-1}\eeq
For $\omega_{0} \gg a$ the equilibrium population of the upper level
relative to the lower level is then
\beq {A_{\uparrow} \over {A_{\downarrow}}} \simeq {1 \over
	{4\sqrt{3}}} {a \over \omega_0 }e^{-2\sqrt{3} {\omega_0
\over a}} \eeq
leading to a temperature
\beq k T_{eff} \simeq {a \over {2 \sqrt{3}}} \eeq
which is higher by a factor ${ 1 \over {\sqrt{3}}} \pi$ than the Unruh
temperature for linear acceleration $T_U = {a \over {2 \pi}}$.

With the help of Eqs. (\ref{eq882}) and (\ref{eq883}), we can determine
from $\langle dH_A /{d\tau}
\rangle_{vf}$ and $\langle dH_A /{d\tau} \rangle_{rr}$ the
contributions of vacuum fluctuations and radiation reaction to the
Lamb shift of the circularly moving two-level atom. We know already that
the contribution of radiation reaction is the same as for an inertial
atom and thus does not contribute to the relative energy shift:
\beq \Delta_{rr} \equiv (\delta E_+)_{rr}-(\delta E_-)_{rr}=0.\eeq
The contribution of vacuum fluctuations can be calculated with
(\ref{eq882}) and (\ref{eq7}):
\beq \left( \delta E_a \right) _{vf} = {\mu^2 \over {8\pi^2}} \sum_b
	|\langle a|R_2^f(0)|b\rangle|^2 \int_0^{+\infty} \,d\omega'
	\left( \omega' +{a \over {2 \sqrt{3}}}Ae^{-2 \sqrt{3} B
	{\omega'\over a}} \right)\left( {{\cal P} \over \omega' +
	\omega_{ab}}-{{\cal P} \over \omega' - \omega_{ab}} \right).
	\marke{eq10}\eeq
The Lamb shift $\Delta$, i. e. the relative shift of the two levels
can be found by evaluating the $b$ summation in (\ref{eq10}) for
each of the levels separately:
\bea \Delta &=& \Delta_{vf} \equiv (\delta E_+)_{vf}-(\delta
E_-)_{vf}\nn\\
	&=& {\mu^2 \over {16\pi^2}}
	\int_0^{+\infty} \,d\omega' \left[ \omega' +{a \over {2
\sqrt{3}}}A
	e^{-2 \sqrt{3} B {\omega'\over a}} \right]\left( {{\cal P} \over
	\omega' + \omega_{0}}-{{\cal P} \over \omega' - \omega_{0}}
	\right). \marke{eq11}\eea
The first one of the two terms appearing in square brackets,
coincides with
the Lamb shift $\Delta_{inert}$ for an atom at rest. The second one
represents
the influence of the circular acceleration, which
 gives the finite correction ${\cal D}$ to the level shift. With
the help of
Eq. (3.354.3) from Ref. \cite{Gradshteyn}, it can be evaluated in
closed form. We obtain
\beq {\cal D} = {a\mu^2 \/64\sqrt{3}\pi^2} \left( e^{-2\sqrt{3}B
	\omega_{0}/a} \, \overline{\hbox{Ei}}(2\sqrt{3}B \omega_{0}/a)
	-e^{2\sqrt{3}B \omega_{0}/a} \, \overline{\hbox{Ei}}
	(-2\sqrt{3}B \omega_{0}/a)\right). \eeq
$\overline{\hbox{Ei}}$ denotes
the principal value of the exponential integral function
\cite{Gradshteyn}.
A plot of the acceleration-induced correction ${\cal D}$ to the inertial
emission rate $\Gamma_{inert}={\mu^2 \over {8 \pi}} \omega_0$ as a
function
of ${a \over {\omega_{0}}}$
is shown in Fig. 1.

\begin{figure}[t]
\epsffile{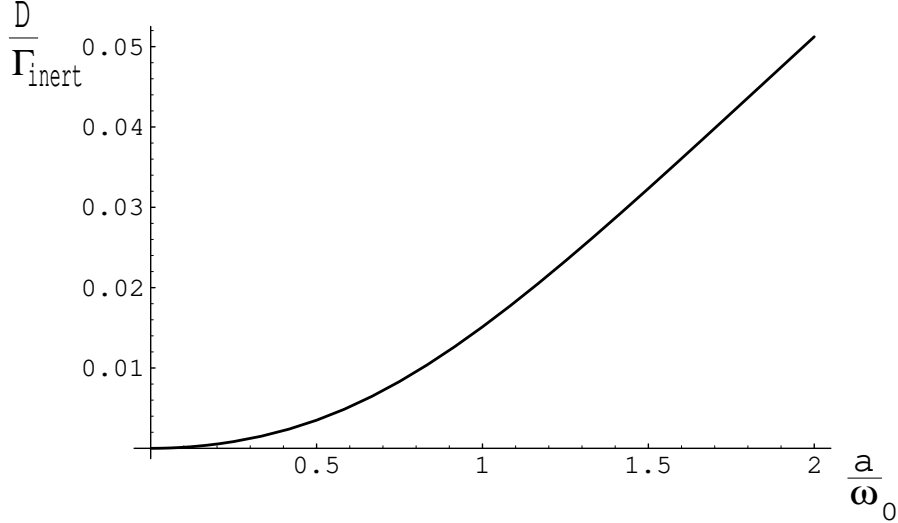}
\caption{The acceleration-induced energy shift ${\cal D}$ of a two-level
atom in circular motion ($\Gamma_{inert}={\mu^2 \over {8 \pi}}
\omega_0$).}

\end{figure}

For an estimation of the order of magnitude we consider an
electron, moving
in a circular orbit perpendicular to a uniform magnetic field
$|{\bf B}_0|=B_0$ (cf. \cite{Bell83}).
The magnetic and electric fields in the co-moving frame are
\beq {\bf B}_0'=\gamma {\bf B}_0, \eeq
\beq {\bf E}_0'=\gamma {\bf v} \times {\bf B}_0 \eeq
The splitting of the two energy levels "spin up" and "spin down" in the
rest frame is given by
\beq \omega_0={e \over {2 m}} |g| B_0' .\eeq
The proper acceleration of an electron moving with velocity v is
\beq a={e \over m} E_0'=v {e \over m} B_0'. \eeq
Therefore we get for ultra-relativistic electrons ($v \simeq 1$,
g=2) the ratio
$a/\omega_0 \approx 1$, which gives rise to a correction ${\cal D}$
of the
relative energy shift with value ${{\cal D} / {\Gamma_{inert}}}
\approx 1.5 \%$.


\end{document}